\documentclass{article}
\usepackage[letterpaper]{geometry}

\usepackage{authblk}
\usepackage{amssymb}
\usepackage{upgreek}
\usepackage{microtype}
\usepackage{enumitem}
\setlist{noitemsep}

\usepackage{color}
\definecolor{keywordcolor}{rgb}{0.7, 0.1, 0.1}   
\definecolor{tacticcolor}{rgb}{0.1, 0.2, 0.6}    
\definecolor{commentcolor}{rgb}{0.4, 0.4, 0.4}   
\definecolor{symbolcolor}{rgb}{0.0, 0.1, 0.6}    
\definecolor{sortcolor}{rgb}{0.1, 0.5, 0.1}      
\definecolor{attributecolor}{rgb}{0.7, 0.1, 0.1} 

\usepackage[colorlinks=true,linkcolor=blue,citecolor=blue,urlcolor=blue]{hyperref}

\usepackage[utf8x]{inputenc}

\usepackage{listings}

\newcommand{\filelink}[1]{\href{https://github.com/starkware-libs/formal-proofs/tree/master/src/#1}{\lstinline{#1}}}

\title{A Verified Algebraic Representation of \\Cairo Program Execution}

\author[1]{Jeremy Avigad}
\author[2]{Lior Goldberg}
\author[2]{David Levit}
\author{Yoav Seginer}
\author[2]{Alon Titelman}
\affil[1]{Carnegie Mellon University}
\affil[2]{StarkWare Industries Ltd.}

\begin{document}

\maketitle

\begin{abstract}
  Cryptographic interactive proof systems provide an
  efficient and scalable means of verifying the results of computation on blockchain.
  A prover constructs a proof, off-chain, that the execution of a program
  on a given input terminates with a certain result.
  The prover then publishes a certificate that can be verified efficiently
  and reliably modulo commonly accepted cryptographic assumptions.
  The method relies on an algebraic encoding of execution traces of programs.
  Here we report on a verification of the correctness of such an encoding of the
  Cairo model of computation with respect to the STARK interactive proof system,
  using the Lean 3 proof assistant.
\end{abstract}

\section{Introduction}
\label{section:introduction}

The execution of \emph{smart contracts} \cite{szabo:97} on blockchain
makes it possible to carry out transactions
such as cryptocurrency exchanges and auctions in the absence of institutional
oversight. But the technology faces a scaling problem: distributed blockchain
protocols require anyone verifying the integrity of the blockchain to carry
out the computation associated with the contract, constituting a draw on resources.

Seminal work of Babai, Fortnow, and Lund \cite{babai:et:al:91} showed
that \emph{interactive proof protocols}, introduced to complexity theory in
the mid-1980s \cite{babai:85,goldwasser:et:al:89}, allow for
efficient verification of computational claims.
They provide protocols by which a computationally powerful \emph{prover}
can convince a computationally limited \emph{verifier} that
the execution of a program on a given input yields a certain result,
without requiring the verifier to execute the program itself.
The original protocols rely on a source of randomness to convince
the verifier that the claim holds with high probability.
Contemporary variants replace the use of randomness by the use of
cryptographic methods and assumptions.

Here we focus on the use of one particular cryptographic protocol,
the \emph{STARK} protocol \cite{ben:sasson:et:al:18},
to verify claims about execution in one particular Turing-complete model of computation,
the \emph{Cairo} machine \cite{goldberg:et:al:21}.
The Cairo model of computation is a machine with a small instruction set
and the ability to read from a bank of memory consisting of values in a finite field.
Roughly speaking, the STARK protocol allows a prover to efficiently
convince a verifier that the prover possesses a table of values from a large finite
field, such that tuples from the table satisfy a family of polynomials that is
shared by both the prover and verifier. In our case, this table of values encodes the trace of an execution of the Cairo machine.

To employ the STARK protocol, claims about the execution of a Cairo program on a
given partial assignment to memory
are encoded using an \emph{algebraic intermediate representation} (AIR).
Specifically, the claim that a program terminates successfully
is expressed as a claim about the existence of a solution
to a family of polynomials, which is exactly the type of claim that can be verified by the
STARK protocol.

When the execution of a smart contract can have substantial financial
repercussions,
it is important to have prior confidence that the outcome
will be as intended.
In that respect, there are at least three aspects of the process
just described that are open to formal verification.
First, one may wish to verify the claimed properties of STARKs,
that is, the claim that the protocol establishes the existence of the
data satisfying the polynomial constraints with high probability,
under the relevant cryptographic assumptions.
Second, one may wish to verify the algebraic representation of
program execution,
that is, the claim that the existence of data satisfying the AIR
implies the existence of an execution trace of the corresponding program.
Finally, one may wish to verify claims that \emph{particular} Cairo
programs meet their specifications, for example,
that the successful execution of a program correctly determines the outcome of an exchange.

Although the first task would constitute an interesting formalization
project, we take the other two tasks to be more pressing.
Cryptographic protocols have been well-studied
and the papers on STARKs and related
protocols have appeared in peer-reviewed journals,
so it seems reasonable to treat the protocol as a black box for now.
Work on the third task is in progress, but we will not report on that here.
This paper reports on the successful completion of the second task,
a fully verified proof that data satisfying the AIR implies
the corresponding computational claim. The AIR verified by the proof is directly
generated from the Cairo implementation, allowing for continuous verification
of the AIR used in production.

When formal verification and computational complexity come together,
the phrase ``interactive proof system'' is used in two unrelated ways.
In formal verification, it refers to a computational proof assistant,
whereas in complexity theory and cryptography,
it refers to the kind of interactive proof protocols described above.
The terminology is standard in both communities, and it is generally not hard
to resolve the ambiguity in context.

Our formalization follows the informal proofs---some presented in detail,
others sketched---in what we will refer to as the
\emph{Cairo whitepaper} \cite{goldberg:et:al:21}.
The most up-to-date version of the formalization, as well as instructions for
compiling or browsing it in Lean, can be found at
\begin{quote}
  \href{https://github.com/starkware-libs/formal-proofs}{https://github.com/starkware-libs/formal-proofs}.
\end{quote}
In the PDF version of this paper, all \texttt{.lean} file references are hyperlinked to the
files in this repository.

\section{The Formalization Platform}

Our formalization is carried out in the \emph{Lean 3} proof assistant \cite{de:moura:et:al:15}
with its associated library, \emph{mathlib} \cite{mathlib}.
Lean's axiomatic foundation is a version of dependent type theory with inductive types
and a type of propositions.
The core logic is constructive, but we (and mathlib) make free use of
classical logic.

Our formalization does not depend heavily on the specifics of Lean and mathlib, however.
We use basic facts from the library about natural numbers, integers,
finite indexing types, bitvectors, and finite fields. Perhaps the most mathematically
involved fact we need is that a degree $n$ polynomial over a field has at most $n$ roots.
We also make moderate use of automation.
Lean's simplifier, which does conditional term rewriting with a battery of tagged
and/or explicitly enumerated rewrite rules, has been helpful,
especially for the parts of the formalization described in Section~\ref{section:one:step:correctness}.
We use a tactic, \emph{norm-num}, that carries out verified numeric calculations.

\section{The Cairo CPU}
\label{section:the:cairo:cpu}

The successful run of a Cairo program on a given input is meant to
convince a skeptical verifier of some claim about those inputs.
An unusual feature of the model of computation is that memory is
\emph{read only}.
When executing a Cairo program and preparing the data needed
for the interactive proof protocol,
it is the prover's obligation to arrange data in memory
so that the program executes successfully.

The computation model is based on a CPU with three registers,
a \emph{program counter} (pc), an \emph{allocation pointer} (ap), and a
\emph{frame pointer} (fp), each of which points to locations in the read-only memory.
The program counter contains the memory address of the instruction
that is about to be executed.
By convention, the allocation pointer points to a yet-unused memory cell,
and it usually only increases as a program executes.
The frame pointer is used to point to a function's local memory.
When a function is called, the frame pointer is set equal to the allocation
pointer, and when the function returns,
the frame pointer is restored to the value it had before the previous call,
that is, the address of the calling function's local memory.

In the Cairo model, register values, as well as values in memory,
are elements of a finite field, $F$.
As a result, there is no order on the elements;
a program can test equality of elements but cannot compare them to
determine which is greater.
Using a field means that we can add, subtract, multiply, and divide.
Program instructions are 63 bits long, and so can be represented by
an integer less than $2^{63}$.
The semantics assumes (and the algebraic data presented to the verifier
guarantees) that the characteristic of the underlying field is greater than
$2^{63}$, which means that instructions have a unique representation
in the field.
But algebraic constraints are needed to guarantee that a given field element
represents an instruction in such a way, as well as to reason about
the instruction's components.
We will explain how this works in Section~\ref{section:instructions}.

The fact that memory is read only means that we do not need to
verify claims about memory management.
Another consequence of the intended use is that we do not need to prove that a
program terminates;
the algebraic data verified by the interactive proof system guarantees
that the program has run for a certain number of steps, after which the program
counter reached a certain value.
That is not to
say that it would not be helpful to also have general guarantees that
high-level Cairo programs terminate and manage memory appropriately;
such \emph{completeness} claims guarantee that the prover can
publish the desired proofs.
At present, extensive testing and code review is used for that purpose,
and establishing \emph{soundness}, i.e.~the fact that the proofs do what they
are supposed to, is the more pressing concern.

There are a few basic types of Cairo machine instructions:
\begin{itemize}
  \item \emph{assert} statements, which assert equality between two values
  \item \emph{conditional and unconditional jumps}, the former based on a test for zero
  \item \emph{call} and \emph{return}
  \item an instruction to advance the allocation pointer.
\end{itemize}
Arguments can refer to the contents of memory locations
using offsets from the frame pointer or allocation pointer.
They also support addition and multiplication.
Each instruction contains three bitvectors of length 16,
which can provide, for example, memory offsets for the operands and result
of an \emph{assert}. Each instruction also contains 15 1-bit flags,
which determine things like the
type of instruction, the nature of the operands, and whether the allocation
pointer should be augmented after the instruction.

An instruction is formally represented in Lean as follows:
\begin{lstlisting}
  structure instruction :=
  (off_dst : bitvec 16)
  (off_op0 : bitvec 16)
  (off_op1 : bitvec 16)
  (flags   : bitvec 15)
\end{lstlisting}
For example, in the Cairo assembly language, we might write an instruction
as follows:
\begin{verbatim}
  [ap + 10] = [fp] * [fp - 1].
\end{verbatim}
This asserts that the value in memory at location {\tt ap + 10} is
the product of the values at locations {\tt fp} and {\tt fp - 1},
where {\tt ap} and {\tt fp} denote the allocation pointer and frame
pointer, respectively.
This instruction is encoded in the structure above by setting the flags
to specify that the operation is an assert, that the relevant operation
is multiplication, that the operands are addressed by offset
from {\tt fp}, and that the result is addressed by offset from {\tt fp}.
The values of \verb~off_dst~, \verb~off_op0~, and \verb~off_op1~
represent 10, 0, and -1, respectively.
The instruction is ultimately converted to a number, which, in turn,
is stored in memory as an element of $F$.
Some instructions also require an immediate value, which is stored
in the next element of memory.

The register state is represented formally as follows:
\begin{lstlisting}
  structure register_state (F : Type*) :=
  (pc : F) (ap : F) (fp : F)
\end{lstlisting}
A single state of the virtual Cairo machine is given by the contents
of memory, which we model as a function from $F$ to $F$,
and the current register state.
Given such a state, the Cairo whitepaper \cite{goldberg:et:al:21}
specifies the potential successor states.
The successful execution of a well-formed instruction leads to a
unique successor state. But there may not be a successor state,
for example, when an assertion fails because in the given state
the asserted equality doesn't hold. And ill-formed instructions
result in undefined behavior, giving rise to the possibility
of multiple successor states.
We therefore use a next-state relation to model the step semantics.
The file \filelink{cpu.lean}, less than 200 lines, is
a straightforward formalization of the whitepaper description,
culminating in the following definitions:
\begin{lstlisting}
  def instruction.next_state (i : instruction) (mem : F → F)
    (s t : register_state F) : Prop :=
  (i.next_pc mem s).agrees t.pc ∧
  (i.next_ap mem s).agrees t.ap ∧
  (i.next_fp mem s).agrees t.fp ∧
  i.asserts mem s

  def next_state (mem : F → F)
    (s t : register_state F) : Prop :=
  ∃ i : instruction, mem s.pc = ↑i.to_nat ∧ i.next_state mem s t
\end{lstlisting}
The first definition says that \lstinline{t} is a successor
state to \lstinline{s} assuming \lstinline{i} is the current instruction.
It relies on auxiliary definitions that specify the next program counter,
the next allocation pointer, the next frame pointer, and any
assertions associated with the instruction.
Notation such as \lstinline{i.next_pc}
relies on a nice bit of Lean syntax known as ``anonymous projections'':
given \lstinline{i} of type \lstinline{instruction} and a
definition \lstinline{instruction.next_pc} in the environment,
Lean interprets \lstinline{i.next_pc mem s} as
\lstinline{instruction.next_pc i mem s}. (In this example, \lstinline{i} is
inserted as the first argument of type \lstinline{instruction}.)
Each of the functions \lstinline{next_pc}, \lstinline{next_ap},
and \lstinline{next_fp} returns an element of an option type,
equal to \lstinline{none} if the instruction results in undefined behavior.
If \lstinline{x} is an element of an option type,
the relation \lstinline{x.agrees a} says that if \lstinline{x} is of the form
\lstinline{some b}, then \lstinline{a} is equal to \lstinline{b}.

The second definition, the \lstinline{next_state} relation,
simply asserts that the memory at the program counter contains
the cast of an instruction to the field and that the next
state agrees with the one corresponding to that instruction.

Other definitions specify the way that
the value of the next state is determined based on the instruction flags
and offset values.
For example, the next value of the frame pointer depends on whether
the current instruction is a call, a return, an assert, or
none of the above.
The specification has the following shape,
where the auxiliary function \lstinline{instruction.dst mem s} determines
the appropriate
value for restoring the frame pointer on the return from a function call.
\begin{lstlisting}
  def next_fp : option F :=
    match i.opcode_call, i.opcode_ret,
      i.opcode_assert_eq with
    | ff, ff, ff := some s.fp
    | tt, ff, ff := some (s.ap + 2)
    | ff, tt, ff := some (i.dst mem s)
    | ff, ff, tt := some s.fp
    | _,  _,  _  := none
    end
\end{lstlisting}
The function is a case distinction on the settings of the corresponding
flags of the instruction.
After a call instruction, for example, the frame pointer is set
to the current allocation pointer plus 2.
After a return instruction, the next value of the frame pointer is
computed using the function \lstinline{instruction.dst}.
The other instructions do not change the frame pointer.

The Cairo toolchain allows programmers to write Cairo programs in
an assembly language that is close to the machine code, with
instructions like these:
\begin{verbatim}
  [ap] = [fp + (-4)] + [fp + (-3)]; ap++
  jmp rel 4 if [ap + (-1)] != 0
  call rel -16
\end{verbatim}
The first instruction says that the
value of the memory location referenced by the allocation pointer
is equal to the sum of the values at two locations
referenced using the frame pointer.
The \lstinline{ap++} suffix specifies that the
instruction also increments the allocation pointer.
The next instruction is a conditional relative jump,
and the third executes a call to the function at the
current program counter minus 16.

The Cairo toolchain also allows programmers to write programs in a higher-level
language with function definitions, data structures,
conditional blocks, and recursive function calls.
These are compiled to assembly instructions,
which are encoded as machine instructions and ultimately as field elements
that are shared with the verifier.

\section{The AIR}
\label{section:the:air}

In the intended scenario, the prover and the verifier
agree on a Cairo program of interest and the input to that Cairo program.
The prover wants to convince the verifier that executing the program on the inputs yields
a certain output.
The prover does this by publishing a \emph{proof}, a suitable certificate,
at which point the verifier executes an algorithm that uses the certificate to
ascertain the correctness of the claim.
The catch is that, in the intended usage, the certificate will be published
on blockchain and the verifier will be executed as part of a smart contract,
so we want the certificate to be small and the verification to be substantially
more efficient than executing the Cairo program itself.

Whereas the original interactive proof systems made use of randomness to achieve such goals,
blockchain protocols rely on the use of cryptographic hash functions instead of coin flips.
The effectiveness of the proof relies on the assumption that a dishonest but
computationally-bounded prover does not have a substantially better than random chance of
gaming the use of a hash.

Our method relies on the STARK protocol, which works roughly as follows.
Fix a large prime number $p$, and let $F_p$ be the finite field of integers modulo $p$.
The STARK protocol allows the prover to
convince the verifier that the prover is in possession of a
two-dimensional table of field elements of $F_p$
that satisfies a constraint system that the prover and verifier agree on.
The constraint system is given as a list of polynomials
$P_1(\vec x_1), \ldots, P_s(\vec x_s)$ over $F_p$,
and a corresponding list of domains $D_1, \ldots, D_s$,
which are periodic subsets of the row indices.
The arguments of the polynomials can be taken from multiple rows and multiple columns
of the table.
We say that the constraint system is satisfied by a certain
table with $N$ rows if
for every $j=1,\ldots,s$ and every $r \in D_j$,
we have $P_j(\vec x_j^{(r)}) = 0$,
where $\vec x_j^{(r)}$ is $\vec x_j$ shifted by $r$ rows.
For example, a polynomial $P(\vec x)$ might involve variables in two consecutive rows,
and the corresponding set $D$ might specify that the constraint holds
at every row in the table, or that it holds at every other row,
or that it holds at every fourth row.
In our application, the polynomials are fixed modulo some field parameters that the
prover and verifier share. The verification algorithm runs
in time polynomial in the logarithm of $N$ and the size of the parameters
that the prover and verifier share, and it provides high assurance that the
result is correct.

To meet our goals, it therefore
suffices to design a sequence of polynomials $P_1(\vec x_1), \ldots, P_s(\vec x_s)$
such that, for parameters corresponding to the Cairo program, input data, and output
in question, and for a suitable $N$,
the existence of a table of data meeting the criteria
described in the previous paragraph implies the existence of an execution sequence of the given Cairo program
on the given input data, yielding the given output.

For the purposes of this paper, the details of the STARK protocol and the precise
complexity and probabilistic claims can be treated as a black box.
Our goal here is to describe the polynomials $P_1, \ldots, P_s$ and to describe
a formal proof that the existence of the table of data implies
the existence of the desired execution sequence.
Assuming the verifier trusts the STARK protocol, the successful run of the
verification algorithm yields a strong guarantee of the existence of
a table of data satisfying the constraints,
and our formal proof turns this into a strong guarantee that the Cairo
program executes as claimed.

The constraints $P_1, \ldots, P_s$ and the relevant parameters
are listed in \filelink{constraints_autogen.lean}.
As the name suggests, this file is automatically generated.
In fact, it is generated by the same code
that produces the code for the verifier.
This, together with the specification of the CPU semantics in \filelink{cpu.lean},
are the only two files that the statement of our final theorem depends on,
other than general basic facts about data structures, fields, and so on in the library.
The desired conclusion, namely, the existence of an execution trace consistent
with the given partial assignment to memory, relies on the fact that the
STARK certificate that is verified on the blockchain is correct with respect to these
polynomial constraints and parameters.
A skeptic can inspect the smart contract that performs the verification
to ensure that this is the case.

The fact that the Cairo program, the input, and the output are all stored in memory
allows us to simplify the description of the verification task.
It suffices for the prover and verifier to agree on a partial assignment of
values to the memory that includes that data and the initial state of the CPU.
In fact, we share that information, the length of the execution trace, and the final value of the
program counter and allocation pointer; the polynomial constraints ensure that the initial value
of the frame pointer is equal to the initial value of the allocation pointer.
This data is represented as follows:
\begin{lstlisting}
  structure input_data (F : Type*) :=
  (trace_length : nat)
  (initial_ap : nat) (initial_pc : nat)
  (final_ap : nat) (final_pc : nat)
  (m_star : F → option F)
\end{lstlisting}
The name \lstinline{m_star} corresponds to the fact that the partial assignment
is written as $\mathfrak m^*$ in the whitepaper, and the return type, \lstinline{option F}, means
that the function can either return an element of the type \lstinline{some a},
where \lstinline{a} is an
element of \lstinline{F}, or \lstinline{none}.

The publicly shared parameters also include additional data,
including information about \emph{range checked} elements and the \emph{interaction} elements;
these are explained in
Sections~\ref{section:permutations}--\ref{section:memory} below.
\begin{lstlisting}
  structure public_data (F : Type*) :=
  (memory_perm_interaction_elm : F)
  (memory_hash_interaction_elm0 : F)
  (memory_public_memory_prod : F)
  (rc16_perm_interaction_elm : F)
  (rc16_perm_public_memory_prod : F)
  (rc_min : nat) (rc_max : nat)
  (initial_rc_addr : nat)
\end{lstlisting}
The correctness proof assumes that this data satisfies certain assumptions
that can be easily confirmed by the verifier.
\begin{lstlisting}
  structure public_constraints
    (inp : input_data F)
    (pd : public_data F) : Prop :=
  (rc_max_lt : pd.rc_max < 2^16)
  (trace_length_le_char : inp.trace_length ≤ ring_char F)
  ...
\end{lstlisting}

Suppose the Cairo program terminates after $T$ steps.
The execution trace then consists of $T + 1$ register states,
including the start state and end state,
and depends on $T$-many instructions in the memory along the way.
Without loss of generality, we can assume that $T + 1$ is a power of two,
since, by convention, Cairo programs ``terminate'' by entering an infinite loop.
The data in the Algebraic Intermediate Representation (AIR) consists of a list
of $16 (T + 1)$-many tuples, each consisting of 25 columns.
These are used to encode the execution trace,
the instructions, and the contents of memory, in ways that we will describe
below.
The parameter \lstinline{trace_length} in the structure \lstinline{input_data}
indicated above is actually $16 (T + 1)$,
corresponding to the parameter $N$ described earlier
in this section.
We recover $T$ by dividing by 16 and subtracting one.
We often refer to the elements of the AIR as \emph{trace cells}, since they encode
the states of the execution trace, as well as memory accesses along the way and auxiliary data.

We can now state our main result:
our formal proof shows that satisfaction of the polynomial constraints
guarantees that the encodings have the claimed meaning with respect to the formal semantics.
In other words, the existence of a table of data satisfying the contraints
implies the existence of an execution trace that is consistent with the agreed-upon
partial assignment to memory.
For the protocol, it is important that the prover commit to the first 23 columns
of data before ``random'' interaction elements are computed
based on a cryptographic hash of the first columns;
the remaining 2 columns then depend on the result of these interaction elements.
From the published data the verifier can ensure that this is, in fact, the case,
and in the next section we will see how they are incorporated into the correctness proof.

The Cairo whitepaper describes the relevant polynomials and encoding of data
without specifying exactly how the data is laid out in the $16 (T + 1) \times 25$ array.
This makes the constraints easier to read and understand.
Our initial formalization followed the whitepaper;
the file \filelink{constraints.lean} formalizes those constraints, and the file
\filelink{correctness.lean} proves the correctness theorem in those terms.
We then incorporated a file \filelink{glue.lean} that mediates between the
two representations,
resulting in the end-to-end correctness proof in \filelink{final_correctness.lean}.

\section{The Correctness Theorem}
\label{section:the:correctness:theorem}

The statement of the final correctness theorem depends only on
the specification of the Cairo execution semantics in \filelink{cpu.lean}
and the autogenerated file of constraints, \filelink{constraints_autogen.lean}.
The full statement of the theorem is presented in Figure~\ref{fig:final:correctness}.

In the statement of the theorem, \lstinline{F} is assumed to be a finite field
of characteristic at least $2^{63}$. We are given the shared input data,
\lstinline{inp : input_data F}, and the remaining public data, \lstinline{pd : public_data F},
assumed to satisfy the required constraints.
We are also given \lstinline{c: columns F}, a structure that consists of
23 columns named \lstinline{c.column0} to \lstinline{c.column22}.

Based on this data, we assert the existence of three ``bad'' subsets
\lstinline{bad1}, \lstinline{bad2}, \lstinline{bad3} of the field $F$.
Letting $T'$ be the shared input parameter \lstinline{inp.trace_length},
these are asserted to have cardinality at most $(T' / 2)^2$, $T'/ 2$, and $T'$,
respectively.
The idea is that for the execution sequences that arise in practice,
these values will be substantially smaller than the cardinality of $F$,
so that a randomly chosen element of the field is unlikely to land in the
bad sets.
As we explain in Section~\ref{section:permutations},
if the prover has encoded the relevant data honestly, these sets are all empty,
but if the prover is dishonest, the prover will be caught in the lie unless the
interaction elements happen to return values in these small sets.
Once these bad sets are determined, the prover commits to the final two columns,
\lstinline{ci : columns_inter F}.

The theorem then says that if the columns \lstinline{c} and \lstinline{ci}
satisfy all the polynomial constraints,
and assuming that the generated interaction elements are not in the small bad sets,
then there exists an execution trace of the Cairo program
meeting the agreed-upon specification.

In greater detail, we assume that the data \lstinline{c}, \lstinline{ci},
\lstinline{inp}, and \lstinline{pd} satisfies the autogenerated constraints
\begin{itemize}
\item \lstinline{cpu__decode},
\item \lstinline{cpu__operands},
\item \lstinline{cpu__update_registers},
\item \lstinline{cpu__opcodes},
\item \lstinline{memory},
\item \lstinline{rc16},
\item \lstinline{public_memory}, and
\item \lstinline{initial_and_final}.
\end{itemize}
We will describe the contents of these below.
We also assume that the interaction elements have landed outside the small bad sets
that depend only on the first 23 columns. Under the cryptographic assumptions, this occurs with high probability.
Let $T$ be $(T' - 1) / 16$.
The theorem then asserts that there exists an assignment to the memory
that extends the agreed-upon partial assignment
and an execution trace of the Cairo program, for that memory assignment,
that starts at the start state, runs for $T$ steps,
and ends at the end state.
In other words, the first and last state of the execution trace
are as claimed, and each successive state follows the previous one
according to the machine semantics specified in \filelink{cpu.lean}.

\begin{figure}[ht]
\begin{lstlisting}
theorem final_correctness
  {F : Type} [field F] [fintype F]
  (char_ge : ring_char F ≥ 2^63)
  /- public data -/
  (inp : input_data F)
  (pd  : public_data F)
  (pc  : public_constraints inp pd)
  (c   : columns F) :
  /- sets to avoid -/
∃ bad1 bad2 bad3 : finset F,
  bad1.card ≤ (inp.trace_length / 2)^2 ∧
  bad2.card ≤ inp.trace_length / 2 ∧
  bad3.card ≤ inp.trace_length ∧
∀ ci : columns_inter F,
  /- autogenerated constraints-/
  cpu__decode c ∧
  cpu__operands c ∧
  cpu__update_registers c inp ∧
  cpu__opcodes c ∧
  memory inp pd c ci ∧
  rc16 inp pd c ci ∧
  public_memory c ∧
  initial_and_final inp c ∧
  /- probabilistic constraints -/
  pd.hash_interaction_elm0 ∉ bad1 ∧
  pd.interaction_elm ∉ bad2 ∧
  pd.interaction_elm ≠ 0 ∧
  pd.rc16__perm__interaction_elm ∉ bad3 →
  /- the conclusion -/
  let T := inp.trace_length / 16 - 1 in
  ∃ mem : F → F,
    option.fn_extends mem inp.m_star ∧
    ∃ exec : fin (T + 1) → register_state F,
      (exec 0).pc = inp.initial_pc ∧
      (exec 0).ap = inp.initial_ap ∧
      (exec 0).fp = inp.initial_ap ∧
      (exec (fin.last T)).pc = inp.final_pc ∧
      (exec (fin.last T)).ap = inp.final_ap ∧
      ∀ i : fin T,
        next_state mem (exec i.cast_succ)
          (exec i.succ)
\end{lstlisting}
\caption{The final correctness theorem.}
\label{fig:final:correctness}
\end{figure}

What needs to be trusted?
The polynomials that are found in
\filelink{constraints_autogen.lean}
should match the polynomials used by the verifier.
Since the definition of a field, the definition of integers, finite sequences,
and so on come into the definition of the semantics and the statement of
correctness, one needs to assume that they have been formalized correctly,
so that, for example, a statement about a finite field really is a statement
about a finite field.
Since the formalization says something about the CPU semantics,
we have to accept that the formalization of that semantics matches our
informal understanding. However, if we are later able
to prove that programs meet high level specifications with respect to the semantics
(which we intend to do, as described in Section~\ref{section:introduction}),
at that point only the higher-level specification matters;
the execution semantics becomes only a stepping-stone to the final result.
Of course, we have to trust the soundness of the axiomatic foundation
and its implementation in Lean.
Finally, as described in Section~\ref{section:introduction},
an implementation of the method also requires
trusting the STARK protocol and the means used to verify
the STARK certificates.

Once we trust the formalization, we do not need to
worry about why or how the polynomial constraints guarantee the existence
of the final execution trace.
The formal proof establishes that it does.
From the point of view of the design of the system, however,
and from the point of view of verification, this is the
most interesting part.
The details are spelled out in the Cairo whitepaper \cite{goldberg:et:al:21}.
In the sections that follow we sketch the proof
and provide some
indications as to how it is formalized.

\section{Summary of the Constraints}
\label{section:constraints}

Altogether, the Cairo algebraic intermediate representation of an execution trace is encoded by the following constraints:
\begin{itemize}
  \item constraints for specifying and decoding instructions;
  \item constraints that specify the next-step relation, and, in particular,
    specify the operands for an assert statement and the values of the program counter,
    frame pointer, and allocation pointer, all in terms of the current instruction and values
    in memory;
  \item constraints that show that the prover has an assignment of values to memory that is
    consistent with program execution and extends the partial assignment that the prover
    and verifier have agreed upon; and
  \item constraints that guarantee that memory addresses and instruction components are
    integers in the required range.
\end{itemize}
We describe each of these categories, in turn, in the sections that follow.

The full list of constraints is found in Section 9.10 of the Cairo whitepaper \cite{goldberg:et:al:21}. They are stated as local hypotheses in the files throughout the
formalization, and they are gathered into structures in the file \filelink{constraints.lean}.
That file provides the parameters and hypotheses that are used in the statement of
the correctness theorem
in \filelink{correctness.lean}. The file \filelink{constraints_autogen.lean} contains a description of the constraints that is generated automatically from the Cairo
source code,
and so reflects the data that is used to generate the STARK certificate.
It is these latter constraints that the verifier can check using the STARK protocol.
We therefore use the file \filelink{glue.lean} to instantiate
the structures and hypotheses in \filelink{constraints.lean}, and the final correctness theorem
in Figure~\ref{fig:final:correctness} is stated in terms of the autogenerated constraints.
This means that the verifier does not have to trust any aspect of the implementation:
the STARK certificate ensures that the constraints are met, and the formal theorem shows
that this implies the existence of the claimed execution trace.

\section{Decoding Instructions}
\label{section:instructions}

In Section~\ref{section:one:step:correctness}, we will discuss polynomial constraints that,
given an instruction $i$, the contents of memory $\mathfrak m$, and
a register state $s$, specify that another register state $t$
follows $s$ according to $i$.
For that purpose, however, we need to reason about the components of $i$.
We also need to ensure that the field element $i$ is, in fact, a valid
instruction, which is to say, it is the result of casting an integer in $[0, 2^{63})$
to the finite field.
Note that we can always use auxiliary elements of the AIR when writing
constraints. These are existentially quantified by the STARK protocol.

Remember that an instruction contains three 16-bit numbers,
$\mathrm{off}_{\tt dst}$, $\mathrm{off}_{\tt op0}$, and $\mathrm{off}_{\tt op1}$.
These are intended to denote integer offsets in $[-2^{15}, 2^{15})$,
but in the AIR we store the corresponding elements
$\widetilde{\mathrm{off}}_* = \mathrm{off}_* + 2^{15}$
(where $*$ is one of {\tt op0}, {\tt op1}, or {\tt dst}).
We also need to reason about the flags $f_0, \ldots, f_{14}$.
Rather than using trace elements of the AIR for these,
we define $\tilde f_i = \sum^{14}_{j=i} 2^{j - i} \cdot f_j$ and store
$\tilde f_0, \ldots, \tilde f_{15}$.
Note that this means that $\tilde f_0 = \sum_{j < 15} 2^j \cdot f_j$
is the integer in $[0, 2^{15})$ with bits $f_0, \ldots, f_{14}$, and that $\tilde f_{15} = 0$.
We can recover the bits $f_i$ with the identity $f_i = \tilde f_i - 2 \tilde f_{i + 1}$.

In Section~\ref{section:range:checks} we explain how the AIR can constrain
the values $\widetilde{\mathrm{off}}_*$ to be integers in the range $[0, 2^{16})$.
Setting that aside, the AIR should therefore assert that elements $\tilde f_i$ encode
a sequence of 0s and 1s via the identity above, and that the value
representing the instruction is equal to
$\widetilde{\mathrm{off}}_{\tt dst} +
2^{16} \cdot \widetilde{\mathrm{off}}_{\tt op0} +
2^{32} \cdot \widetilde{\mathrm{off}}_{\tt op1} +
2^{48} \cdot \tilde f_0$.
These constraints are written in Lean as follows:
\begin{lstlisting}
  variable h_instruction : inst = off_dst_tilde + 2^16 * off_op0_tilde +
    2^32 * off_op1_tilde + 2^48 * f_tilde 0

  variable h_bit : ∀ i : fin 15, f_tilde.to_f i * (f_tilde.to_f i - 1) = 0

  variable h_last_value : f_tilde ⟨15, by norm_num⟩ = 0
\end{lstlisting}
To make sense of this, note that in Lean, a hypothesis is encoded as a variable whose type
is the statement in question. The type \lstinline{fin 15} consists of the numbers from 0 to 14,
which are used to index the bits. In contrast, the variables $\tilde f_i$ are indexed
by elements of \lstinline{fin 16}, and the Lean idiom \lstinline{⟨15, by norm_num⟩} is
used to denote the last element of this type and automatically fill in the proof that 15 is less
than 16. The \lstinline{to_f} function in the expression \lstinline{f_tilde.to_f i}
computes the value of $f_i$ from the tuple $\tilde f$ via the identity above.

Theorem 1 of the whitepaper asserts that with these constraints (and the ones
that ensure that the values $\widetilde{\mathrm{off}}_*$ are integers in the right range)
there is a unique instruction such that the field element $\mathrm{inst}$ encodes
that instruction.
In our formalization, we generally need to refer to \emph{the} instruction encoded by
$\mathrm{inst}$ on the assumption that these constraints are met.
Therefore, our formalization takes the following form.
First, we define a function
\begin{lstlisting}
  def the_instruction : instruction := ...
\end{lstlisting}
This specifies the instruction from the data $\widetilde{\mathrm{off}}_*$ and $\tilde f$.
(In Lean, when variables and hypotheses are declared in a file, definitions that use them
can leave the dependence implicit. The dependencies are part of the definition, however,
and are displayed when the user asks Lean to show an expression or its type.)
Then, we prove that, assuming the constraints are met, the field element is equal to
the cast (to the field) of the natural number encoding of the instruction, and the trace elements
$\widetilde{\mathrm{off}}_*$ are the results of casting the components of the
instruction to the field. For example:
\begin{lstlisting}
  theorem inst_eq : inst = ↑(the_instruction ...).to_nat

  theorem off_dst_tilde_eq : off_dst_tilde =
    ↑(the_instruction ...).off_dst.to_natr
\end{lstlisting}
Here, we have omitted the arguments to \lstinline{the_instruction} and the proofs of the theorems.
The up arrow, which indicates the cast, can often be left implicit.
The first theorem says that if we take the instruction computed from the
constraints, encode it as a natural number less than $2^{63}$ in
the natural way, and cast it to the finite field,
the result is exactly \lstinline{inst}.
The second theorem says, similarly,
that the AIR element \lstinline{off_dst_tilde}
is the field element corresponding to the \lstinline{off_dst} bitvector
component of the instruction.
The function \lstinline{to_natr} translates the given bitvector to a natural number;
the ``r'' at the end reflects the fact that the \lstinline{to_nat} function in
Lean's standard library
takes $0$ to be the most significant bit rather than the least significant bit,
whereas \lstinline{to_natr} does the opposite.
We also prove uniqueness:
\begin{lstlisting}
  theorem inst_unique (i : instruction) (h : inst = i.to_nat) :
    i = the_instruction ...
\end{lstlisting}
Interestingly, uniqueness is not required for the correctness theorem in
Figure~\ref{fig:final:correctness}.
The possibility that a field element might represent different instructions would
make the next-state relation nondeterministic, but it would not violate the theorem.
But it is independently important to know that this nondeterminism does not arise,
and we use that fact in subsequent work, when we want to prove that particular programs
meet their specifications.
For that purpose, we need to know that the original instruction can be recovered from its
representation in the field.

\section{The Next State Relation}
\label{section:one:step:correctness}

The constraints that relate one state to the next are rather straightforward.
The AIR has cells associated to each step $i \le T$ of the
execution trace.
There are values $\mathit{ap}_i$, $\mathit{fp}_i$, and $\mathit{pc}_i$
representing the allocation pointer, frame pointer, and program counter at step $i$.
There are also
auxiliary values $\mathit{dst}_i$ and $\mathit{res}_i$ for destination and result
values associated with some of the instructions, as described by the CPU specification,
and auxiliary values used in the constraints, as described by the Cairo whitepaper.
Some of the auxiliary values are used to express the constraints as quadratic
expressions, as required by the protocol.

For example, to say that the field value \lstinline{next_fp} represents the
correct value of the next frame pointer given the current register values, the
state of memory, and the current instruction, the AIR uses the following constraint:
\begin{lstlisting}
  variable h_next_fp : next_fp =
    f_tilde.f_opcode_ret * dst + f_tilde.f_opcode_call * (ap + 2) +
      (1 - f_tilde.f_opcode_ret - f_tilde.f_opcode_call) * fp
\end{lstlisting}
\lstinline{f_tilde.f_opcode_ret} and \lstinline{f_tilde.f_opcode_call}
are the values of the corresponding flags, which were decoded by the instruction
constraints. If the \lstinline{opcode_ret} bit is 1 and the
\lstinline{opcode_call} bit is 0, the constraint requires that the next
frame pointer has the value \lstinline{dst};
if \lstinline{opcode_ret} is 0 and \lstinline{opcode_call} is 1,
the constraint requires that the next frame pointer has the value \lstinline{ap} + 2;
if both are $0$, the constraint requires that the frame pointer remains unchanged.
Remember that if, for example, both flags are 1, the CPU semantics
says that the next frame pointer can be anything at all; it is nondeterministic
in that sense.
This is not problematic, because the program itself is part of the specification
shared by the prover and the verifier. The verifier can check that the
instructions are indeed well-formed, and an end-to-end verification that a
Cairo program meets its specification will generally prove that fact
along the way.

The claim that this constraint together with the constraints governing
instructions, calculation of $\mathit{dst}$, memory access, and so on all
imply that the calculated next state agrees with the CPU semantic specification
is stated simply as follows:
\begin{lstlisting}
  theorem next_fp_agrees : (inst.next_fp mem ⟨pc, ap, fp⟩).agrees next_fp
\end{lstlisting}
The parameters and hypotheses are left implicit, which is a good thing,
because there are a lot of them, even for a simple fact like this one.
Asking Lean to display the theorem statement shows that there are
five hypotheses involving eight field values, the memory assignment,
the instruction, and the associated tuple of flags.
As one might expect, one of the most useful roles that the proof assistant
played in our formalization was keeping track of the large array
of data and assumptions,
and making sure all the intermediate results were glued together correctly.
The proof of this particular theorem involving little more than casing
on the values of the flags,
invoking previous results, and simplifying, with about eight lines of tactics.

\section{Permutations}
\label{section:permutations}

There are two types of constraints left to describe.
Defining the next-step relation requires, in certain places, statements to the
effect that a memory location $a$ contains a value $v$. The prover needs
not only to encode this information in the AIR but also to convince the verifier that there
is an assignment of values to the read-only memory that is consistent with these
claims. The constraints that do this are described in Section~\ref{section:memory}.
The correctness of the execution trace also requires knowing that address offsets
stored in an instruction are integers in the interval $[0, 2^{16})$, and
the Cairo compiler allows programmers to make similar assertions in their code.
These \emph{range checks} are handled by constraints described in Section~\ref{section:range:checks}.
Both the memory constraints and range check constraints rely on an additional
feature of the interactive proof protocol, the interaction step,
that was alluded to in
Section~\ref{section:the:air}. In this section we explain how it works.

Suppose the prover has a sequence $a_0, a_1, \ldots, a_{n-1}$ of values
in the field,
uniformly spaced in the data, so that the prover can express uniform claims
about these values using polynomial constraints.
Remember that in general the verifier will not see this data;
the prover will merely use the interactive proof system
to convince the verifier of its existence.

Now suppose that to establish further claims about the data, the prover needs
to list the elements in a different order.
The prover adds a permutation $a'_0, a'_1, \ldots, a'_{n-1}$,
with additional polynomial constraints expressed in terms of those elements.

To establish correctness, the prover needs to convince the verifier that
the second sequence is a permutation of the first.
This is not easy to do.
A STARK makes it possible to write down a constraint that relates,
say, each pair $(a_i, a'_i)$ to the next pair $(a_{i+1}, a'_{i+1})$,
but the statement that the sequence $(a_i)_{i < n}$ is a reordering of
$(a'_i)_{i < n}$ cannot be expressed using local constraints of that form.

This is where the interaction step comes in. Let $p(x)$ be the polynomial given by
$p(x) = \prod_{i < n} (x - a_i)$, and let $p'(x) = \prod_{i < n} (x - a'_i)$.
If the sequence $(a_i)_{i < n}$ is a permutation of $(a'_i)_{i < n}$, then $p(x) - p'(x)$
is the zero polynomial.
If it isn't, then $p(x) - p'(x)$ is a polynomial of degree at most $n$,
and so has at most $n$ roots in the field.
Suppose $n$ is much smaller than the size of the field, and let $\alpha$ be a
randomly chosen element.
Suppose also that the prover convinces the verifier that $p(\alpha) - p'(\alpha) = 0$.
Then the verifier knows that either $(a_i)_{i < n}$ really is a permutation of $(a'_i)_{i < n}$,
or the prover was very lucky that the choice of $\alpha$ did not disprove that claim.

The interactive proof protocol uses a cryptographic hash instead of randomness.
The STARK protocol enables the prover to commit to $a_0, a_1, \ldots, a_{n-1}$
and $a'_0, a'_1, \ldots, a'_{n-1}$ and publish a certificate that these values
meet the publicly shared constraints.
Once it is published, a cryptographic hash $\alpha$ is generated.
The prover then publishes a certificate that establishes that
$p(\alpha) - p'(\alpha) = 0$.

To implement the argument in Lean, given any pair of finite tuples \lstinline{a}
and \lstinline{b} indexed over a finite type \lstinline{n}, we define the set of
misleading choices of $\alpha$ as follows:
\begin{lstlisting}
  variables {F : Type*} [field F] [fintype F]
  variables {n : Type*} [fintype n]
  variables (a b : n → F)

  def exceptional_set : finset F :=
    if ∀ z, ∏ i, (z - a i) = ∏ i, (z - b i) then ∅
    else univ.filter (λ z : F, ∏ i : n, (z - a i) = ∏ i, (z - b i))
\end{lstlisting}
If the two polynomials are equal, the exceptional set is the empty set, which is to say, there
are no misleading values.
Otherwise, the exceptional set is the finite set of values that make the two sides equal.
The facts we need to establish are,
first, that the cardinality of the exceptional set is less than that of
\lstinline{n},
and, second, that unless the field element \lstinline{z} is in the
exceptional set, equality of the two polynomials at \lstinline{z} implies
that they are always equal.
\begin{lstlisting}
  theorem card_exceptional_set_le : card (exceptional_set a b) ≤ fintype.card n

  theorem all_eq_of_not_mem_exceptional_set {z : F}
      (h₁ : z ∉ exceptional_set a b)
      (h₂ : ∏ i : n, (z - a i) = ∏ i, (z - b i)) :
    ∀ z, ∏ i, (z - a i) = ∏ i, (z - b i)
\end{lstlisting}
The first claim is proved by cases on the definition of the
exceptional set. If it is the empty set, the claim is trivial,
and otherwise the claim follows from the fact that the
elements of the exceptional set are the roots of a nonzero
polynomial of degree at most the cardinality of \lstinline{n}.
The second claim is immediate from the definition of
the exceptional set,
since the hypotheses imply that it cannot be equal to the
second branch of the conditional.

In the proof of the final theorem in Figure~\ref{fig:final:correctness},
the existential quantifiers over \lstinline{bad2} and \lstinline{bad3}
are witnessed by
instances of \lstinline{exceptional_set}. (We take \lstinline{bad1} to be the
set of zeros of another polynomial,
described in Section~\ref{section:memory}.)
Note that the statement of the final theorem doesn't depend on the
definition of \lstinline{exceptional_set} or the details of how the
bad sets are constructed.
The theorem simply asserts the existence of finite sets that depend on the first
23 columns of data and not the final two,
with the properties that (1) they are small, and (2) as long as the
values of the interaction elements are outside those sets,
the desired conclusion is guaranteed to hold.

\section{Range Checks}
\label{section:range:checks}

We have seen that specifying that a trace cell represents a well-formed instruction
requires in particular showing that certain other trace cells are field elements that
are casts of integers in the interval $[0, 2^{16})$.
Suppose $a_0, a_1, \ldots, a_{n-1}$ are the relevant values, let $\mathrm{rc}_\mathrm{min}$ be
the minimum value, and let $\mathrm{rc}_\mathrm{max}$ be the maximum value.
We can, without loss of generality, assume that every value between $\mathrm{rc}_\mathrm{min}$ and
$\mathrm{rc}_\mathrm{max}$ occurs on the list, by padding the list with extra elements if necessary.
We can convince the verifier that all the values are between $\mathrm{rc}_\mathrm{min}$ and $\mathrm{rc}_\mathrm{max}$ by dedicating another sequence of trace values $a'_0, a'_1, \ldots, a'_{n-1}$,
making $\mathrm{rc}_\mathrm{min}$ and $\mathrm{rc}_\mathrm{max}$ available to the verifier,
and establishing the following claims:
\begin{itemize}
  \item The sequence $(a'_i)_{i < n}$ is a permutation of $(a_i)_{i < n}$.
  \item For each $i < {n-1}$, either $a_{i + 1} = a_i$ or $a_{i + 1} = a_i + 1$.
  \item $a'_0 = \mathrm{rc}_\mathrm{min}$ and $a'_{n-1} = \mathrm{rc}_\mathrm{max}$.
\end{itemize}

The first of these is handled as described in Section~\ref{section:permutations}.
We designate a sequence of trace elements $p_0, p_1, \ldots, p_{n-1}$ with the declaration
\lstinline{p : fin (n + 1) → F}, and we add the following constraints:
\begin{lstlisting}
  (z - a' 0) * p 0 = z - a 0
  ∀ i : fin n, (z - a' i.succ) * p i.succ =
    (z - a i.succ) * p i.cast_succ
  p (fin.last n) = 1
\end{lstlisting}
Here \lstinline{i.succ} is Lean notation for $i + 1$ as a value of type \lstinline{fin (n + 1)},
and \lstinline{i.cast_succ} is notation for $i$ as a value of type \lstinline{fin (n + 1)}.
These constraints guarantee that we have $\prod_{i < n} (z - a_i) = \prod_{i < n} (z - a'_i)$, and
if $z$ is generated by a hash, this equation offers the verifier a strong guarantee that the
first condition is met.
As described in Section~\ref{section:permutations}, given
\begin{lstlisting}
  def bad_set_3 (a a' : fin (n + 1) → F) : finset F :=
  polynomial_aux.exceptional_set a a'
\end{lstlisting}
and the hypothesis \lstinline{z ∉ bad_set_3 a a'}, we have
\begin{lstlisting}
  lemma rc_permutation : ∀ i, ∃ j, a i = a' j
\end{lstlisting}

For the second and third conditions, we simply add the following constraints:
\begin{lstlisting}
  ∀ i : fin n, (a' i.succ - a' i.cast_succ) *
    (a' i.succ - a' i.cast_succ - 1) = 0
  a' 0 = rc_min
  a' (fin.last n) = rc_max
\end{lstlisting}
With the ultimate arrangement of data in the AIR, the values \lstinline{off_op0_tilde}
are included along the values \lstinline{a j} with an explicit embedding:
\begin{lstlisting}
  ∀ i, a (embed_off_op0 i) = off_op0_tilde i
\end{lstlisting}
With this, and the assumption \lstinline{rc_max < 2^16}, we have
\begin{lstlisting}
  theorem off_op0_in_range (i : fin T) :
    ∃ k : ℕ, k < 2^16 ∧ off_op0_tilde i = ↑k
\end{lstlisting}
We also have the analogous properties for \lstinline{off_op0_tilde} and \lstinline{off_dst_tilde},
which were assumed in Section~\ref{section:instructions}.

\section{Memory}
\label{section:memory}

During the course of its execution, a Cairo program will access memory locations
$a_0, a_1, \ldots, a_{n-1}$. The prover has to establish the existence of a
list of suitable values $v_0, v_1, \ldots, v_{n-1}$ for those memory locations.
The constraints described in Section~\ref{section:one:step:correctness} guarantee that
the values chosen by the prover are consistent with the semantics of the program execution.
For example, if a program asserts the equality of the values in memory at locations
$a_i$ and $a_j$, the constraints will guarantee that $v_i$ is equal to $v_j$.

But the prover also has to establish the values are consistent with \emph{each other},
which is to say, there exists an assignment $\mathfrak m$ of values to the memory
with the property that $\mathfrak m(a_i) = v_i$ for every $i < n$.
This is equivalent to saying that for every $i$ and $j$ less than $n$, if
$a_i = a_j$, then $v_i = v_j$.

The constraints in the AIR that ensure this are similar to the range check constraints
discussed in Section~\ref{section:range:checks}, with two additional twists.
As in Section~\ref{section:range:checks}, we can assume that the set of values
$\{ a_0, a_1, \ldots, a_{n-1} \}$ is an interval by assigning $0$ to unused memory locations
between the smallest and largest values.
If we designate trace elements $a'_0, a'_1, \ldots, a'_{n-1}$ and
$v'_0, v'_1, \ldots, v'_{n-1}$,
it suffices to establish the following:
\begin{itemize}
\item The sequence of pairs $(a'_i, v'_i)_{i < n}$ is a permutation of the sequence of pairs
  $(a_i, v_i)_{i < n}$.
\item For each $i < n - 1$, either $a'_{i + 1} = a'_i$ or $a'_{i + 1} = a'_i + 1$.
\item For each $i < n - 1$, if $a'_{i + 1} = a'_i$, then $v'_{i + 1} = v'_i$.
\end{itemize}
The second and third of these are ensured by the following constraints:
\begin{lstlisting}
  ∀ i : fin n, (a' i.succ - a' i.cast_succ) *
    (a' i.succ - a' i.cast_succ - 1) = 0
  ∀ i : fin n, (v' i.succ - v' i.cast_succ) *
    (a' i.succ - a' i.cast_succ - 1) = 0
\end{lstlisting}
Notice that the first of these constraints is needed to establish that the
second one works as advertised.

The first twist is that to establish the first condition, it suffices to establish that the
values $(a'_i + \alpha v'_i)_{i < n}$ are a permutation of the values
$(a_i + \alpha v_i)_{i < n}$, provided that $\alpha$ has the property that $a'_i + \alpha v'_i$
is not equal to $a_i + \alpha v_i$ unless $a'_i = a_i$ and $v'_i = v_i$.
But if the two quantities are equal and $v'_i \ne v_i$, we have
$\alpha = (a'_i - a_i) / (v_i - v'_i)$.
Assuming $n$ is much smaller than the characteristic of $F$, there are a relatively small number
of values $\alpha$ in the field with this property.
So if we have the identity
\[
  \prod_{i < n} (z - (a_i + \alpha v_i)) = \prod_{i < n} (z - (a'_i + \alpha v'_i))
\]
for a randomly chosen $\alpha$ and a randomly chosen $z$, then
first condition holds with high probability.
This explains the expressions \lstinline{bad1} and \lstinline{bad2} in the statement of
the final correctness theorem: the first is the small set of possibly misleading $\alpha$s,
and the second is the small set of possibly misleading $z$s.
These are similar to the set \lstinline{bad3}
defined for the range checks.

The second twist is that our global specification assumes that the prover and the verifier
both have access to a partial specification $\mathfrak m^*$ of values to memory
locations. Remember, this is generally used to specify the Cairo program that has been
executed, the agreed-upon inputs, and the claimed output.
The prover has to convince the verifier that there is a memory function $\mathfrak m$ that
not only maps $a_i$ to $v_i$ for each $i < n$ but is also consistent with
$\mathfrak m^*$.
One solution is to add the set of pairs
$\{ (a, \mathfrak m^*(a)) \}_{a \in \mathrm{dom}(\mathfrak m^*)}$ to the list of pairs
$(a_0, v_0), (a_1, v_1), \ldots, (a_{n-1}, v_{n-1})$,
but that is inefficient, because it requires adding an additional pair of constraints for each element of the domain of $\mathfrak m^*$.
Instead, we add $| \mathrm{dom}(\mathfrak m^*) |$ pairs $(0, 0)$ to the sequence $\ldots (a_i, v_i) \ldots$,
and then we change the constraints to say that the sequence $(a'_i, v'_i)_{i < n}$ is a
permutation of the sequence that results from replacing these pairs $(0, 0)$ by the pairs
$(a, \mathfrak m^*(a))$.
As explained by the whitepaper, it suffices to change the final constraint in Section~\ref{section:range:checks} from
\lstinline{p (fin.last n) = 1} to the following:
\begin{lstlisting}
  p (fin.last n) * ∏ a : mem_dom mem_star, (z - (a.val + alpha * mem_val a)) =
    z^(fintype.card (mem_dom mem_star))
\end{lstlisting}
Assuming all the constraints are met,
we define a function $\mathfrak m$ from the sequences
$(a'_i)_{i < n}$ and $(v'_i)_{i < n}$:
\begin{lstlisting}
  def mem (a' v' : fin (n + 1) → F) : F → F :=
  λ addr, if h : ∃ i, a' i = addr then v' (classical.some h) else 0
\end{lstlisting}
If a value \lstinline{addr} occurs in the sequence \lstinline{a'}, we define \lstinline{mem addr}
to be the corresponding value of \lstinline{v'}. We later show that the constraints imply
that this value is unique.
If \lstinline{addr} does not occur in the sequence, we return 0.

Given this definition of $\mathfrak m$ (i.e.~\lstinline{mem} in the formalization), we then show that the memory constraints imply the following:
\begin{itemize}
  \item The memory assignment \lstinline{mem} extends the partial assignment
    \lstinline{mem_star}, which is ultimately specified in the \lstinline{inp} parameter
    in the final correctness theorem in Figure~\ref{fig:final:correctness}.
  \item The pairs of values \lstinline{a i, v i} referred to in the one-step constraints
    satisfy \lstinline{mem (a i) = v i}.
\end{itemize}
We ultimately have to be explicit as to where the various memory accesses referred to in the cpu
semantics are embedded in the list of pairs $(a_i, v_i)$. This information is specified in \filelink{glue.lean}, which relates the automatically generated polynomial constraints to the informal ones in the
whitepaper description.

This concludes our description of the constraints and our presentation of the main theorem.

\section{Conclusions}
\label{section:conclusions}

We have verified the correctness of an algebraic encoding of the Cairo CPU semantics. The
encoding is used to publish, on blockchain, efficient proofs of the correctness of claims about
machine-code execution with respect to that semantics. The verification is valuable
in its own right, but it is also an important stepping-stone toward verification of
higher-level programs in the Cairo programming language with respect to higher-level
descriptions of their behavior.

A notable feature of our work is that the encoding that we have verified is already in
commercial use.
It is used by the \emph{StarkEx} platform to carry
out cryptocurrency exchanges efficiently, and it is used by \emph{StarkNet},
which enables developers to
implement similarly efficient computational transactions on blockchain.
This places a high premium on establishing the correctness of the encoding.
The whitepaper proof is quite technical, since it requires reasoning about a number
of algebraic constraints and fitting a number of small results together in just the right way.
It is therefore reassuring that the verification of the whitepaper went through straightforwardly
and confirmed the correctness of the implementation.

\section{Related Work}

The formalization we report on here only addresses the relationship between the
machine semantics and its algebraic representation.
It is therefore similar to projects that model processor instruction sets, such as the
x86 instruction set \cite{goel:et:al:20, dasgupta:et:al:19}
and the Ethereum virtual machine \cite{hildenbrandt:18, hirai:17, amani:18}.
In comparison to x86 and even EVM, however, the Cairo machine model is fairly simple.
We have described some of the novel features of the model, such as the fact that it has a
read-only memory and operates on values from a finite field.

We do not know of any project that verifies an algebraic encoding of
execution traces.
Another approach to verifying computation using cryptographic protocols
involves compiling each individual program to a set of polynomial
constraints that describe its execution, unrolling loops and bounding
the number of iterations, and so on. This method suffers from some
well-known drawbacks, discussed in \cite[Section 1.1]{goldberg:et:al:21}.
Fournet et al.~\cite{fournet:et:al:16} verify the correctness of such
transformations, as carried out by Pinocchio \cite{costello:et:al:15,parno:et:al:16}.

The more general goal of verifying smart contracts in various senses
has become too big an industry to survey here \cite{tolmach:et:al:12}.
Jiao et al.~\cite{jiao:et:al} and Ribeiro et al.~\cite{ribeiro:et:al:20}
verify programs written in a subset of Solidity with respect to a high-level description
of the semantics,
and Bhargavan et al.~\cite{bhargavan:et:al:16} verify smart contracts written in Solidity by
translating them to $\mathsf{F}^\star$.
Annenkov et al.~\cite{annenkov:et:al:21} provide means of defining and verifying smart contracts in Coq and then extracting code for various blockchain platforms. A number of systems provide means of
verifying the correctness of cryptographic protocols and their
implementations, including \cite{abate:et:al:21,almeida:et:al:10, almeida:et:al:12,barthe:et:al:10,butler:et:al:21,almeida:et:al:21,sidorenco:et:al:21}.

Our long-term goal of verifying Cairo programs with respect to a machine semantics
(and hence, with the results here, with respect to the final STARK certificates)
makes it similar to other projects that are designed to verify
software with respect to a machine level semantics, such as CompCert \cite{leroy:09},
CakeML \cite{kumar:et:al:14}, and VST \cite{cao:et:al:19}.
Once again, the specifics of the Cairo platform and its applications give rise to novel aspects
of the verification task that we will continue to report on in the future.


\bibliographystyle{plain}
\bibliography{verified_cpu_arxiv}

\begin{thebibliography}{10}

\bibitem{abate:et:al:21}
Carmine Abate, Philipp~G. Haselwarter, Exequiel Rivas, Antoine~Van Muylder,
  Th{\'{e}}o Winterhalter, Catalin Hritcu, Kenji Maillard, and Bas Spitters.
\newblock {SSProve}: {A} foundational framework for modular cryptographic
  proofs in {Coq}.
\newblock pages 1--15, 2021.

\bibitem{almeida:et:al:10}
Jos{\'{e}}~Bacelar Almeida, Endre Bangerter, Manuel Barbosa, Stephan Krenn,
  Ahmad{-}Reza Sadeghi, and Thomas Schneider.
\newblock A certifying compiler for zero-knowledge proofs of knowledge based on
  sigma-protocols.
\newblock In Dimitris Gritzalis, Bart Preneel, and Marianthi Theoharidou,
  editors, {\em European Symposium on Research in Computer Security (ESORICS)
  2010}, pages 151--167. Springer, 2010.

\bibitem{almeida:et:al:12}
Jos{\'{e}}~Bacelar Almeida, Manuel Barbosa, Endre Bangerter, Gilles Barthe,
  Stephan Krenn, and Santiago~Zanella B{\'{e}}guelin.
\newblock Full proof cryptography: verifiable compilation of efficient
  zero-knowledge protocols.
\newblock In Ting Yu, George Danezis, and Virgil~D. Gligor, editors, {\em
  Conference on Computer and Communications Security (CCS) 2012}, pages
  488--500. {ACM}, 2012.

\bibitem{almeida:et:al:21}
Jos{\'{e}}~Bacelar Almeida, Manuel Barbosa, Manuel~L. Correia, Karim Eldefrawy,
  St{\'{e}}phane Graham{-}Lengrand, Hugo Pacheco, and Vitor Pereira.
\newblock Machine-checked {ZKP} for {NP} relations: Formally verified security
  proofs and implementations of mpc-in-the-head.
\newblock In Yongdae Kim, Jong Kim, Giovanni Vigna, and Elaine Shi, editors,
  {\em Computer and Communications Security (CCS) 2021}, pages 2587--2600.
  {ACM}, 2021.

\bibitem{amani:18}
Sidney Amani, Myriam B{\'{e}}gel, Maksym Bortin, and Mark Staples.
\newblock Towards verifying ethereum smart contract bytecode in {Isabelle/HOL}.
\newblock In June Andronick and Amy~P. Felty, editors, {\em Certified Programs
  and Proofs (CPP) 2018}, pages 66--77. {ACM}, 2018.

\bibitem{annenkov:et:al:21}
Danil Annenkov, Mikkel Milo, Jakob~Botsch Nielsen, and Bas Spitters.
\newblock Extracting smart contracts tested and verified in {Coq}.
\newblock In Catalin Hritcu and Andrei Popescu, editors, {\em Certified
  Programs and Proofs (CPP) 2021}, pages 105--121. {ACM}, 2021.

\bibitem{babai:85}
L{\'{a}}szl{\'{o}} Babai.
\newblock Trading group theory for randomness.
\newblock In Robert Sedgewick, editor, {\em {ACM} Symposium on Theory of
  Computing (SToC) 1985}, pages 421--429. {ACM}, 1985.

\bibitem{babai:et:al:91}
L{\'{a}}szl{\'{o}} Babai, Lance Fortnow, and Carsten Lund.
\newblock Non-deterministic exponential time has two-prover interactive
  protocols.
\newblock {\em Comput. Complex.}, 1:3--40, 1991.

\bibitem{barthe:et:al:10}
Gilles Barthe, Daniel Hedin, Santiago~Zanella B{\'{e}}guelin, Benjamin
  Gr{\'{e}}goire, and Sylvain Heraud.
\newblock A machine-checked formalization of sigma-protocols.
\newblock In {\em Computer Security Foundations Symposium (CSF) 2010}, pages
  246--260. {IEEE} Computer Society, 2010.

\bibitem{ben:sasson:et:al:18}
Eli Ben{-}Sasson, Iddo Bentov, Yinon Horesh, and Michael Riabzev.
\newblock Scalable, transparent, and post-quantum secure computational
  integrity.
\newblock {\em {IACR} Cryptol. ePrint Arch.}, 2018:46, 2018.

\bibitem{bhargavan:et:al:16}
Karthikeyan Bhargavan, Antoine Delignat{-}Lavaud, C{\'{e}}dric Fournet, Anitha
  Gollamudi, Georges Gonthier, Nadim Kobeissi, Natalia Kulatova, Aseem Rastogi,
  Thomas Sibut{-}Pinote, Nikhil Swamy, and Santiago~Zanella B{\'{e}}guelin.
\newblock Formal verification of smart contracts: Short paper.
\newblock In Toby~C. Murray and Deian Stefan, editors, {\em Programming
  Languages and Analysis for Security (PLAS) 2016}, pages 91--96. {ACM}, 2016.

\bibitem{butler:et:al:21}
David Butler, Andreas Lochbihler, David Aspinall, and Adri{\`{a}} Gasc{\'{o}}n.
\newblock Formalising $\varsigma$-protocols and commitment schemes using
  crypthol.
\newblock {\em J. Autom. Reason.}, 65(4):521--567, 2021.

\bibitem{cao:et:al:19}
Qinxiang Cao, Lennart Beringer, Samuel Gruetter, Josiah Dodds, and Andrew~W.
  Appel.
\newblock Vst-floyd: {A} separation logic tool to verify correctness of {C}
  programs.
\newblock {\em J. Autom. Reason.}, 61(1-4):367--422, 2018.

\bibitem{costello:et:al:15}
Craig Costello, C{\'{e}}dric Fournet, Jon Howell, Markulf Kohlweiss, Benjamin
  Kreuter, Michael Naehrig, Bryan Parno, and Samee Zahur.
\newblock Geppetto: Versatile verifiable computation.
\newblock In {\em Symposium on Security and Privacy (SP) 2015}, pages 253--270.
  {IEEE} Computer Society, 2015.

\bibitem{dasgupta:et:al:19}
Sandeep Dasgupta, Daejun Park, Theodoros Kasampalis, Vikram~S. Adve, and
  Grigore Rosu.
\newblock A complete formal semantics of x86-64 user-level instruction set
  architecture.
\newblock In Kathryn~S. McKinley and Kathleen Fisher, editors, {\em Programming
  Language Design and Implementation (PLDI) 2019}, pages 1133--1148. {ACM},
  2019.

\bibitem{de:moura:et:al:15}
Leonardo~Mendon{\c{c}}a de~Moura, Soonho Kong, Jeremy Avigad, Floris van Doorn,
  and Jakob von Raumer.
\newblock The lean theorem prover (system description).
\newblock In Amy~P. Felty and Aart Middeldorp, editors, {\em Conference on
  Automated Deduction (CADE) 2015}, pages 378--388. Springer, Berlin, 2015.

\bibitem{fournet:et:al:16}
C{\'{e}}dric Fournet, Chantal Keller, and Vincent Laporte.
\newblock A certified compiler for verifiable computing.
\newblock In {\em Computer Security Foundations Symposium (CSF) 2016}, pages
  268--280. {IEEE} Computer Society, 2016.

\bibitem{goel:et:al:20}
Shilpi Goel, Anna Slobodov{\'{a}}, Rob Sumners, and Sol Swords.
\newblock Verifying x86 instruction implementations.
\newblock In Jasmin Blanchette and Catalin Hritcu, editors, {\em Certified
  Programs and Proofs (CPP) 2020}, pages 47--60. {ACM}, 2020.

\bibitem{goldberg:et:al:21}
Lior Goldberg, Shahar Papini, and Michael Riabzev.
\newblock Cairo --- a turing-complete stark-friendly cpu architecture.
\newblock Cryptology ePrint Archive, Report 2021/1063, 2021.
\newblock \url{https://ia.cr/2021/1063}.

\bibitem{goldwasser:et:al:89}
Shafi Goldwasser, Silvio Micali, and Charles Rackoff.
\newblock The knowledge complexity of interactive proof systems.
\newblock {\em {SIAM} J. Comput.}, 18(1):186--208, 1989.

\bibitem{hildenbrandt:18}
Everett Hildenbrandt, Manasvi Saxena, Nishant Rodrigues, Xiaoran Zhu, Philip
  Daian, Dwight Guth, Brandon~M. Moore, Daejun Park, Yi~Zhang, Andrei
  Stefanescu, and Grigore Rosu.
\newblock {KEVM:} {A} complete formal semantics of the ethereum virtual
  machine.
\newblock In {\em Computer Security Foundations Symposium (CSF) 2018}, pages
  204--217. {IEEE} Computer Society, 2018.

\bibitem{hirai:17}
Yoichi Hirai.
\newblock Defining the ethereum virtual machine for interactive theorem
  provers.
\newblock In Michael Brenner, Kurt Rohloff, Joseph Bonneau, Andrew Miller,
  Peter Y.~A. Ryan, Vanessa Teague, Andrea Bracciali, Massimiliano Sala,
  Federico Pintore, and Markus Jakobsson, editors, {\em Financial Cryptography
  and Data Security (FC) 2017}, pages 520--535. Springer, 2017.

\bibitem{jiao:et:al}
Jiao Jiao, Shuanglong Kan, Shang{-}Wei Lin, David San{\'{a}}n, Yang Liu, and
  Jun Sun.
\newblock Semantic understanding of smart contracts: Executable operational
  semantics of solidity.
\newblock In {\em Security and Privacy (SP) 2020}, pages 1695--1712. {IEEE},
  2020.

\bibitem{kumar:et:al:14}
Ramana Kumar, Magnus~O. Myreen, Michael Norrish, and Scott Owens.
\newblock Cakeml: a verified implementation of {ML}.
\newblock In Suresh Jagannathan and Peter Sewell, editors, {\em Principles of
  Programming Languages (POPL) 2014}, pages 179--192. {ACM}, 2014.

\bibitem{leroy:09}
Xavier Leroy.
\newblock Formal verification of a realistic compiler.
\newblock {\em Commun. {ACM}}, 52(7):107--115, 2009.

\bibitem{mathlib}
The mathlib community.
\newblock The lean mathematical library.
\newblock In Jasmin Blanchette and Catalin Hritcu, editors, {\em Certified
  Programs and Proofs (CPP) 2020}, pages 367--381. {ACM}, 2020.

\bibitem{parno:et:al:16}
Bryan Parno, Jon Howell, Craig Gentry, and Mariana Raykova.
\newblock Pinocchio: nearly practical verifiable computation.
\newblock {\em Commun. {ACM}}, 59(2):103--112, 2016.

\bibitem{ribeiro:et:al:20}
Maria Ribeiro, Pedro Ad{\~{a}}o, and Paulo Mateus.
\newblock Formal verification of ethereum smart contracts using isabelle/hol.
\newblock In Vivek Nigam, Tajana~Ban Kirigin, Carolyn~L. Talcott, Joshua~D.
  Guttman, Stepan~L. Kuznetsov, Boon~Thau Loo, and Mitsuhiro Okada, editors,
  {\em Logic, Language, and Security - Essays Dedicated to Andre Scedrov on the
  Occasion of His 65th Birthday}, pages 71--97. Springer, 2020.

\bibitem{sidorenco:et:al:21}
Nikolaj Sidorenco, Sabine Oechsner, and Bas Spitters.
\newblock Formal security analysis of mpc-in-the-head zero-knowledge protocols.
\newblock In {\em Computer Security Foundations Symposium (CSF) 2021}, pages
  1--14. {IEEE}, 2021.

\bibitem{szabo:97}
Nick Szabo.
\newblock Formalizing and securing relationships on public networks.
\newblock {\em First Monday}, 2(9), 1997.

\bibitem{tolmach:et:al:12}
Palina Tolmach, Yi~Li, Shang-Wei Lin, Yang Liu, and Zengxiang Li.
\newblock A survey of smart contract formal specification and verification.
\newblock {\em ACM Comput. Surv.}, 54(7), 2021.

\end{thebibliography}

\end{document}